\documentclass[%
superscriptaddress,
preprint,
 amsmath,amssymb,
 aps,
 prc
]{revtex4-1}

\usepackage{graphicx}
\usepackage{dcolumn}
\usepackage{bm}
\usepackage{soul,xcolor}
\usepackage{amssymb}
\usepackage{mathrsfs}
\usepackage[pdfencoding=auto,psdextra]{hyperref}
\usepackage{CJK}

\setstcolor{blue}

\begin{document}
\begin{CJK*}{UTF8}{}


\title{Classification of pairing phase transition in the hot nucleus}

\author{Yuhang Gao (\CJKfamily{gbsn}高宇航)}
\affiliation{School of Science, Jiangnan University, Wuxi 214122, China.}

\author{Yanlong Lin (\CJKfamily{gbsn}林彦龙)}
\affiliation{School of Science, Jiangnan University, Wuxi 214122, China.}

\author{Lang Liu (\CJKfamily{gbsn}刘朗)}
\email{liulang@jiangnan.edu.cn}
\affiliation{School of Science, Jiangnan University, Wuxi 214122, China.}

\begin{abstract}

The hot nucleus $^{162}\mathrm{Dy}$ is investigated using covariant density functional theory, where the shell-model-like approach treats the pairing correlation. Lee-Yang's theorem is applied to classify the pairing phase transition by analyzing the distribution of zeros of the partition function in the complex temperature plane. The distribution of zeros of the partition function converges with increasing particle numbers and illustrates the characteristics of the phase transition. In our calculations, we determine the first order of the phase transition near the critical temperature. Different seniority states show the pairing phase transition from a superfluid to a normal phase, ranging from fully paired states to completely unpaired states.

%
%

\end{abstract}

\maketitle
\end{CJK*}


\section{Introduction}
\label{sec:1}
Pairing phase transition in atomic nuclei is essential to understand the properties of hot nuclei~\cite{Bohr1998}. Studying the heat capacity behavior near the critical temperature is a way to investigate the pairing correlations. Experimental work over the past decades has led to the discovery of the S-shaped curve of heat capacity with temperature, based on accurately measuring the level density~\cite{PhysRevC.63.021306,PhysRevLett.83.3150,PhysRevC.63.044309,PhysRevC.68.034311}.Meanwhile, similar S-shaped curves have been investigated in many theoretical models, including the nuclear shell model~\cite{PhysRevC.58.3295,PhysRevLett.87.022501,LANGANKE2005360}, mean field model~\cite{PhysRevLett.85.26,PhysRevC.62.044307,PhysRevC.61.044317,PhysRevC.88.034308,LI2019192} and other models~\cite{PhysRevC.63.044301}. These S-shaped curves can be explained as pairing phase transition from the superfluid phase to the normal phase. Although this pairing phase transition has been extensively investigated, its order still needs to be more conclusive~\cite{PhysRevC.66.024322,BELIC2004381,Moretto_2015}. 

According to Ehrenfest's definition, small systems do not exhibit phase transition~\cite{PhysRevLett.84.3511,10.2307/41134053}. However, Lee and Yang proposed a theorem on the distribution of roots of the grand parition function and predicted wide application even in small system~\cite{PhysRev.87.410}. This theorem has been extended to the canonical ensemble through the analytic continuation of the inverse temperature in the complex plane~\cite{Grossmann1967,Grossmann1969}. In Refs.~\cite{PhysRevLett.84.3511,PhysRevA.64.013611}, it has been proposed a classification scheme for phase transition in finite systems, such as atomic systems, based on the distribution of zeros (DOZ) of the canonical partition function in complex temperature. Being slightly modified, a model for investigating and classifying the pairing phase transition in atomic nuclei systems of two or more particles is developed~\cite{PhysRevC.66.024322}. However, the self-consistency between the two approximation methods proposed in Ref.~\cite{PhysRevC.66.024322} can not be guaranteed. 

In our previous work~\cite{PhysRevC.92.044304,Yan_2021}, thermodynamic properties of even-even nuclei and odd-A nuclei have been studied within the covariant density functional theory (CDFT)~\cite{RING1996193,Meng2021}. The shell-model-like approach (SLAP)~\cite{ZENG19831,PhysRevC.50.1388,Meng2006a} can strictly conserve particle number and accurately handle the blocking effect. The microscopic mechanism of the S-shaped curve is described consistently. In this work, we take $^{162} \mathrm{Dy} $ as an example to investigate the order of pairing phase transition within the CDFT+SLAP model with the classification scheme based on the distribution of zeros of the pairing partition function in the complex temperature plane.

This paper is organized as follows. Sec.$\ref{sec:2}$ makes a simple description of the theoretical framework. The results and discussion are given in Sec.$\ref{sec:3}$. The last section is the summary.

\section{Theoretical framework}
\label{sec:2}

The theoretical framework of point coupling CDFT has been explained in details in Ref.\cite{PhysRevC.82.054319}. A Lagrangian density is the beginning point of CDFT, from which the Dirac equation for nucleons with local scalar $S(\bm{r})$ and vector $V^{\mu}(\bm{r})$ potentials can be deduced as
\begin{equation}
\left[ \gamma_{\mu} (i \partial^{\mu} - V^{\mu}) - (m + S) \right] \psi_{\xi} = 0,
\label{eq:1}
\end{equation}
where
\begin{equation}
S(\bm{r}) = \Sigma_{S},  V(\bm{r}) = \Sigma^{\mu} + \vec{\tau} \cdot \vec{\Sigma}^{\mu}_{TV},
\label{eq:2}
\end{equation}
and $\psi_{\xi}$ is Dirac spinor. 

In the SLAP, the pairing correlation is handled by diagonalizing the following Hamiltonian in the multi-particle configurations (MPCs) space, which is constructed by the single-particle levels acquired by the above CDFT method. 
\begin{align}
  H &= H_{\rm s.p.} + H_{\rm pair} \notag
 \\
    &= \sum\limits_{i}\varepsilon_{i}a^{+}_{i}a_{i} -G\sum\limits^{i\neq j}_{i,j>0} a^{+}_{i}a^{+}_{\bar{i}}a_{\bar{j}}a_{j}, 
\label{eq:3} 
\end{align}
where $\varepsilon_{i}$ is the single-particle energy acquired from the Dirac equation (\ref{eq:1}), $\bar i$ is the time-reversal state of $i$, and $G$ represents constant pairing strength.
For a system which has an even particle number $N = 2n$, the MPCs could be established as follows:
\begin{enumerate}
\item fully paired configurations (seniority $s$ = 0):
\begin{equation}
|c_1\bar{c}_1\cdots c_n\bar{c}_n\rangle=a^+_{c_1}a^+_{\bar{c}_1}\cdots a^+_{c_n}a^+_{\bar{c}_n}|0\rangle;
\label{eq:4} 
\end{equation}
\item configurations with two unpaired particles (seniority $s$ = 2)
\begin{equation}
|i\bar{j}c_1\bar{c}_1\cdots c_{n-1}\bar{c}_{n-1}\rangle=a^+_{i}a^+_{\bar{j}}a^+_{c_1}a^+_{\bar{c}_1}\cdots a^+_{c_{n-1}}a^+_{\bar{c}_{n-1}}|0\rangle\quad\quad(i\ne j);
\label{eq:5} 
\end{equation}
\item configurations with more unpaired particles (seniority $s=4, 6, \ldots$), see, e.g., Refs.~\cite{ZENG19831,Meng2006a}. 
\end{enumerate}

The configurations with energies $E_m-E_0 \leq E_c$ are used to diagonalize the Hamiltonian~(\ref{eq:3}), in which $E_m$ and $E_0$ are the energies of the $m$th configuration and the ground-state configuration, respectively. 

After the diagonalization of the Hamiltonian (\ref{eq:3}), the nuclear many-body wave function is shown as
\begin{align}
|\psi_\beta\rangle =&\sum\limits_{c_{1}\cdots c_{n}}{v_{\beta,\,c_1\cdots c_n}}|c_1\bar{c}_1\cdots c_n\bar{c}_n\rangle  \notag \\
& +\sum\limits_{i,j}{\sum\limits_{c_{1}\cdots c_{n-1}}{v_{\beta(ij),\,c_1\cdots c_{n-1}}}|i\bar{j}c_1\bar{c}_1\cdot\cdot\cdot c_{n-1}\bar{c}_{n-1}\rangle} \notag \\
& + \cdots,
\label{eq:6}
\end{align}
where $\beta = 0$ means the ground state, and $\beta = 1, 2, 3, \ldots$ mean the excited states with the excitation energy $E_{\beta}$. $v_{\beta}$ means the coefficient after diagonalization.
The pairing energy and the pairing gap are defined in Refs.~\cite{Meng2006a,Canto1985,Egido1985,RevModPhys.61.131}

By assuming the hot many-body system is a canonical ensemble~\cite{PhysRevC.76.024319}, the nuclear thermodynamic properties are defined in Ref.~\cite{PhysRevC.92.044304}.
The canonical partition function $Z$ can be written as
\begin{equation}
Z = \sum\limits^{\infty}_{\beta=0}\eta(E_{\beta})\,e^{-E_{\beta}/T}
\label{eq:7}
\end{equation}
where $E_{\beta}$ is the excitation energy which can be obtained from the CDFT + SLAP method, and the corresponding level density $\eta (E_{\beta})$ is taken as $2^s$, i.e., the degeneracy of each state.

In our calculations, the MPC space has 16 single particle levels around the Fermi surfaces and 2, 3, 4, 5 pairs of valence particles for both neutron and proton. It also means the highest seniority can be calculated is 10. The pairing strength in our calculations for neutron $G_{n}$ is fixed to 0.29 $\rm MeV$, and $G_{p}$ for proton is 0.32$\rm MeV$ with an energy cutoff $E_{c} = 45 \rm MeV$.

The classification scheme for phase transition is completely based on Ref.~\cite{PhysRevC.66.024322}. 
Therefore, we will simply summarize the main features of the classification scheme. The scheme relies on the DOZ of the canonical partition function in the complex temperature plane.

First, the inverse complex temperature is defined as:
\begin{equation}
\mathcal{B} = \beta + i\tau,
\label{eq:8}
\end{equation}
where $\beta=1/T$ as usual and $\tau$  means the imaginary part of the inverse complex temperature which is measured in $\mathrm{MeV^{-1}}$ in this work. The zeros of the canonical partition function typically line up on curves through the complex temperature plane. The zeros are denoted by $(\beta_{j},\tau_{j})$ with $j$ = 1 $\ldots$ 4 and $j$ increases with increasing distance from the real axis.

If the first three zeros closest to the real axis have been determined, the average inverse distance between zeros can be calculated as
\begin{equation}
\Phi(\widetilde{\tau}_j)=\dfrac{1}{d_j},
\label{eq:9}
\end{equation}
where $\tilde{\tau}_{j}=(\tau_{j}+\tau_{j+1})/2$, $d_{j}=\sqrt{(\beta_{j+1}-\beta_{j})^{2}+(\tau_{j+1}-\tau_{j})^{2}}$. The function $\Phi$ can then be approximated in the vicinity of the real axis by a power law of $\tau_{j}$ only,
\begin{equation}
\Phi(\tau_j)\propto\tau_j^\alpha,
\label{eq:10}
\end{equation}
then, the quantity of interest, $\alpha$, can be calculated by means of
\begin{equation}
\alpha=\dfrac{\ln\Phi(\tau_3)-\ln\Phi(\tau_2)}{\ln\tau_3-\ln\tau_2}.
\label{eq:11}
\end{equation}
As the definition in Ref.~\cite{PhysRevC.66.024322}. 
When $\alpha$ $<$ 0, the system exhibits a first-order phase transition.

\section{Results and discussion}
\label{sec:3}


\begin{figure}[ht]
\centering
 \includegraphics[scale=0.1]{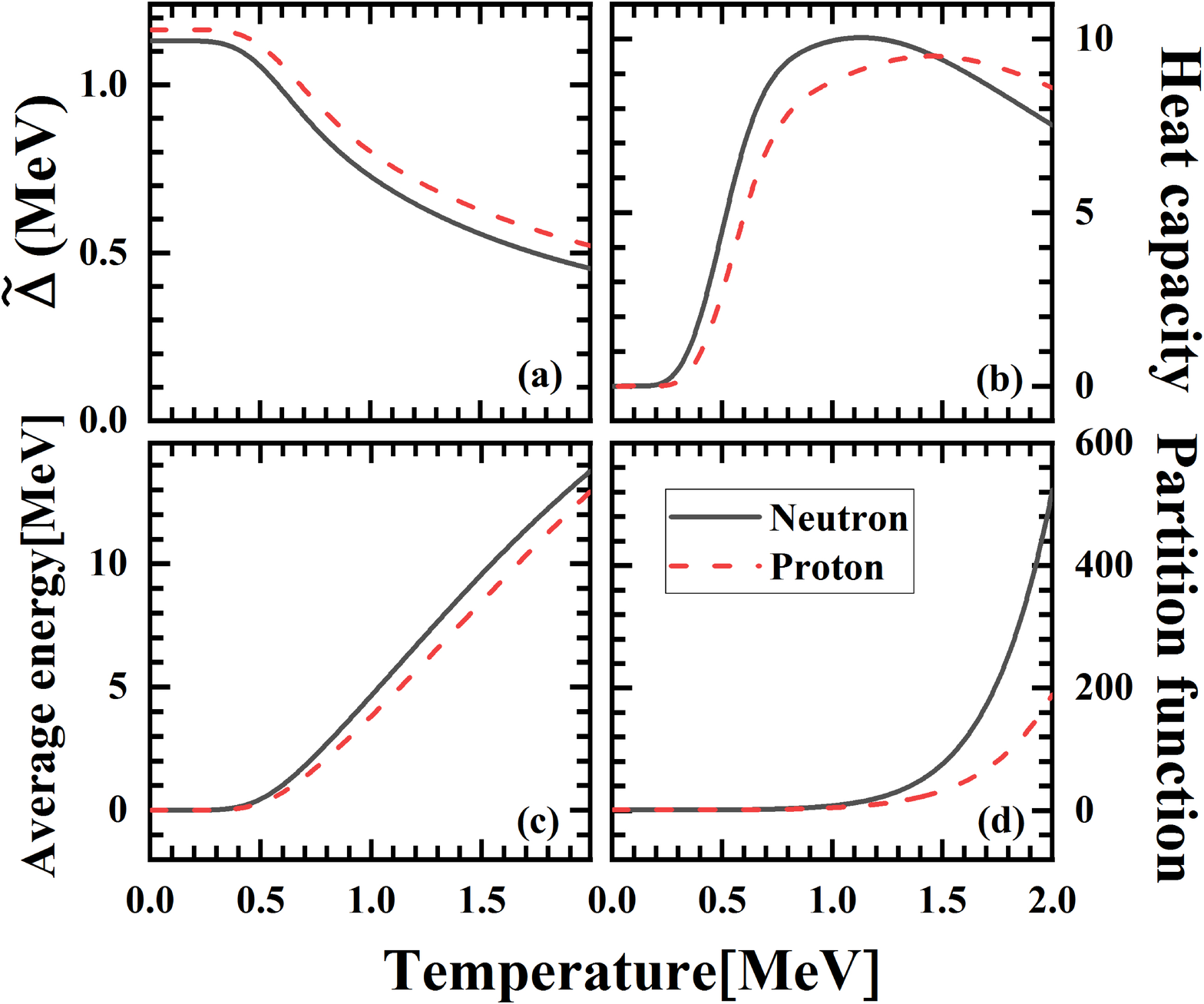}
    \caption{(Color online) The (a) pairing gap, (b) heat capacity, (c) average energy, and (d) partition function of neutron (black solid lines), proton (red dashed line) for $^{162} \mathrm{Dy} $ as functions of temperature.} 
\label{fig:example1}
\end{figure}
The (a) pairing gap, (b) heat capacity, (c) average energy, and (d) partition function have been evaluated in the canonical ensemble theory and shown in Fig.$\ref{fig:example1}$. Because of the particle number conservation, the pairing gap decreases smoothly and does not vanish at temperatures up to 2 $\mathrm{MeV}$. Meanwhile, an apparent S-shape heat capacity curve with respect to the temperature has been shown. The temperature of the second turning point in this curve is usually considered as the critical temperature for the pairing phase transition from the superfluid phase to the normal phase. However, it should be noted that the location of the second turning point in the S-shape curve of the heat capacity could be affected by the limitation of model space. In the meantime, the order of the phase transition of pairing correlations cannot be distinguished directly 
from these results. 
%


\begin{figure}[ht]
\centering
\includegraphics[scale=0.1]{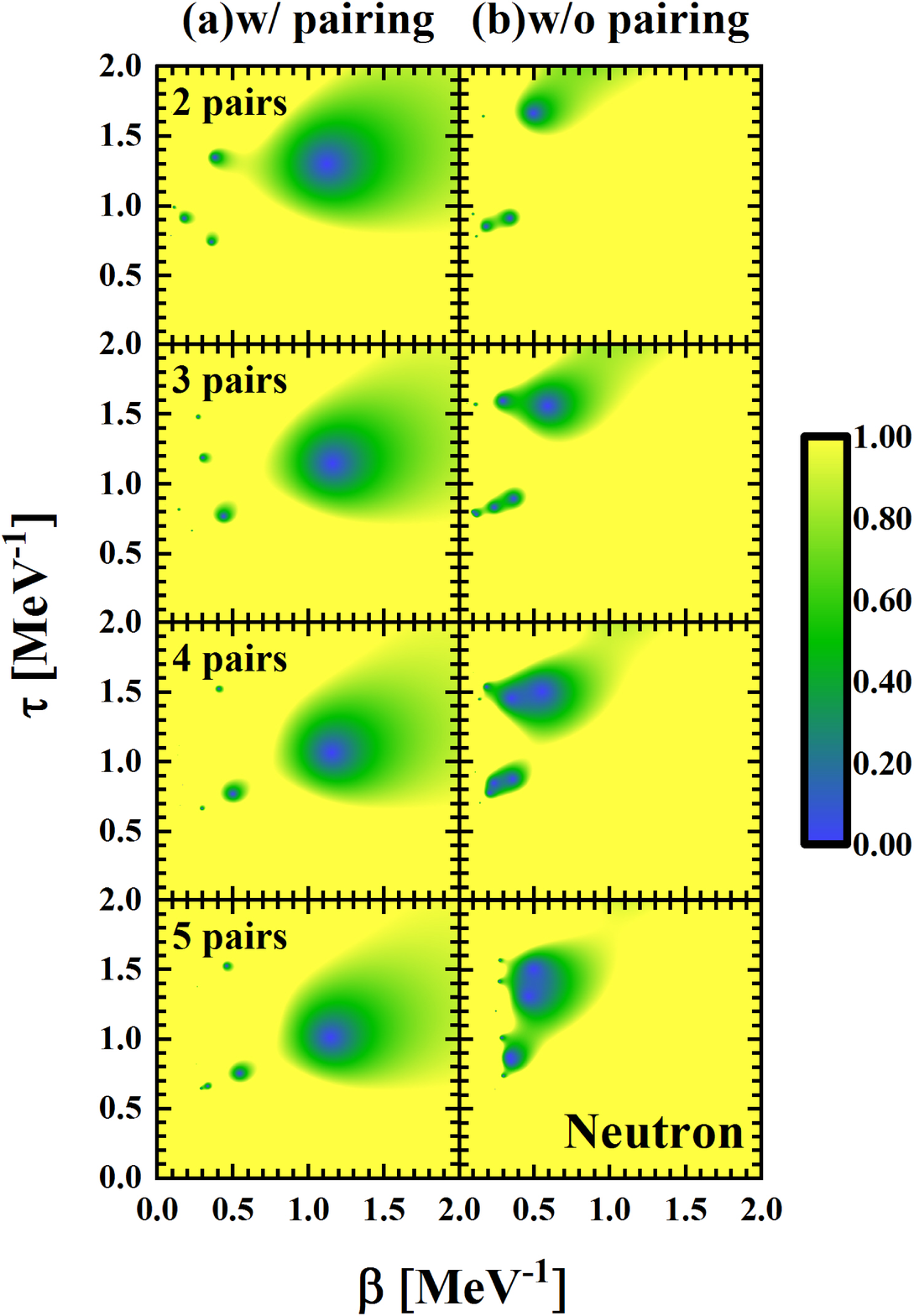}
\caption{(Color online) The contour plots of the partition function of neutron for $^{162} \mathrm{Dy} $ with (a) and without (b) pairing in the complex temperature plane with different pairs of particles.} 
\label{fig:example2}
\end{figure}
Based on the earlier works of Lee and Yang and others~\cite{PhysRevC.66.024322,PhysRevC.66.024322,PhysRev.87.410,Grossmann1967,Grossmann1969,PhysRevA.64.013611}, our calculations can also be extended to the complex temperature plane. Figure $\ref{fig:example2}$ shows the contour plots of the partition function of neutron for $^{162} \mathrm{Dy} $ with (a) and without (b) pairing in the complex temperature plane with different pairs of particles. Different rows represent the DOZ of the partition function for neutron with different pairs from $2$ to $5$. The left and right columns show the cases with and without pairing, respectively. It can be found that there is no zero point in the region at $\beta$ $>$ 1 $\mathrm{MeV^{-1}}$ for the DOZ without pairing. However, the DOZ with pairing shows one zero point at $\beta$ $\approx$ 1.2 $\mathrm{MeV^{-1}}$. It corresponds to the critical point of the $\mathrm{S}$-shape heat capacity curve at $T$ $\approx$ 0.83 $\mathrm{MeV}$ for neutron in Fig.$\ref{fig:example1}$, the critical temperature of the pairing phase transition from the superfluid phase to the normal phase. Furthermore, the DOZ has similar characteristics at $\beta$ $>$ 1 $\mathrm{MeV^{-1}}$ and tends to be same at $\beta$ $<$ 1 $\mathrm{MeV^{-1}}$ with the increasing number of pairs.



\begin{figure}[ht]
\centering
 \includegraphics[scale=0.1]{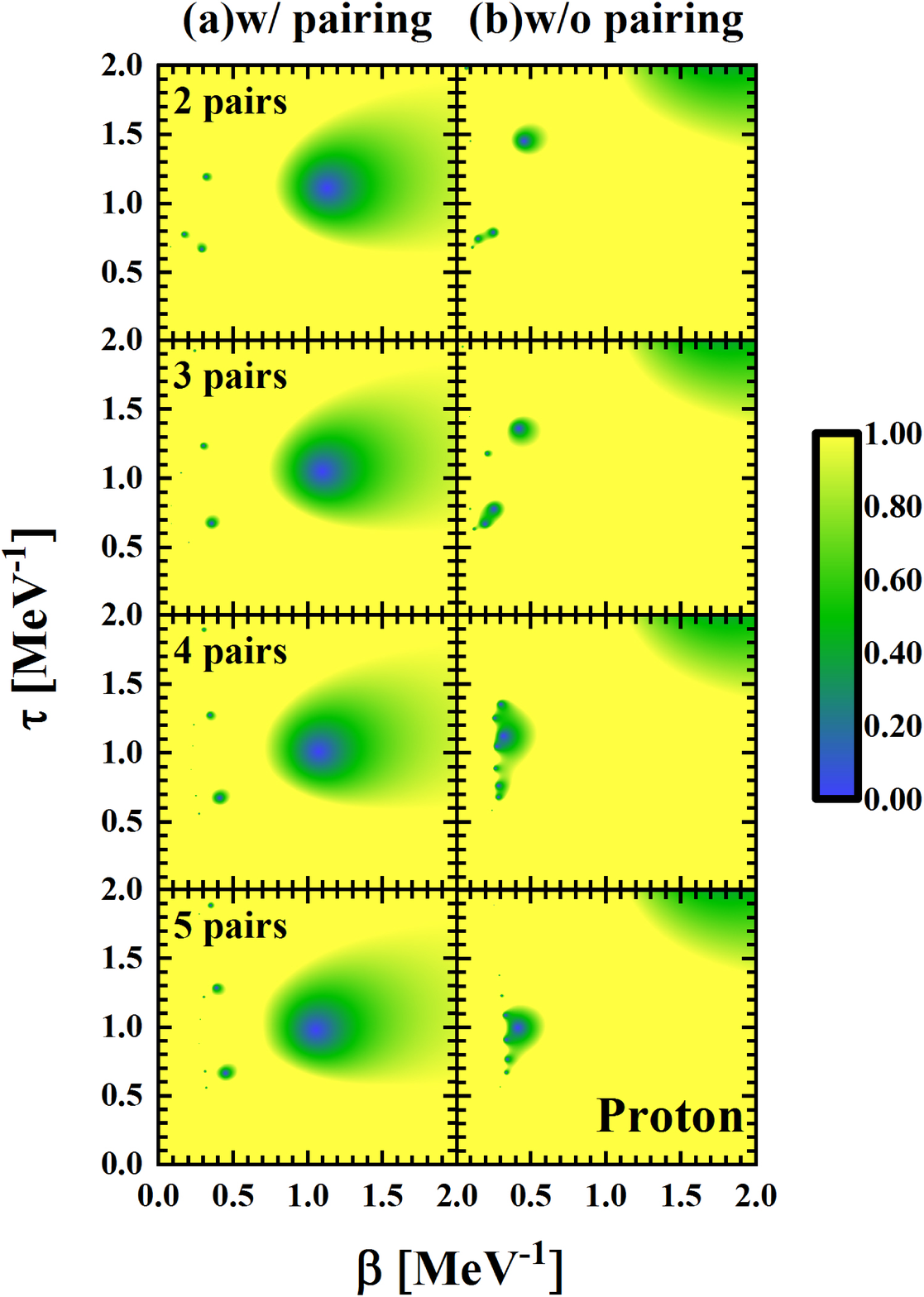}
\caption{(Color online) The contour plots of the partition function of proton for $^{162} \mathrm{Dy} $ with (a) and without (b) pairing in the complex temperature plane  with different pairs of particles.} 
\label{fig:example3}
\end{figure}
The contour plots of the partition function of proton for $^{162} \mathrm{Dy} $ with (a) and without (b) pairing in the complex temperature plane with different pairs of particles are shown in Fig.$\ref{fig:example3}$. Similar to the case of the neutron, there are significant differences between the DOZ with and without pairing, especially in the region at $\beta$ $ > $ 1 $\mathrm{MeV^{-1}}$. The DOZ without pairing shows no zero point, while the DOZ with pairing has a zero point at $\beta$ $\approx$ 1.1 $\mathrm{MeV^{-1}}$, which represents the critical point of the $\mathrm{S}$-shaped heat capacity curve at $T$ $\approx$ 0.90 $\mathrm{MeV}$ for proton in Fig.$\ref{fig:example1}$. The DOZ with pairing converges at $\beta$ $ < $ 1 $\mathrm{MeV^{-1}}$ as the number of pairs increasing.



\begin{figure}[ht]
\centering
    \includegraphics[scale=0.1]{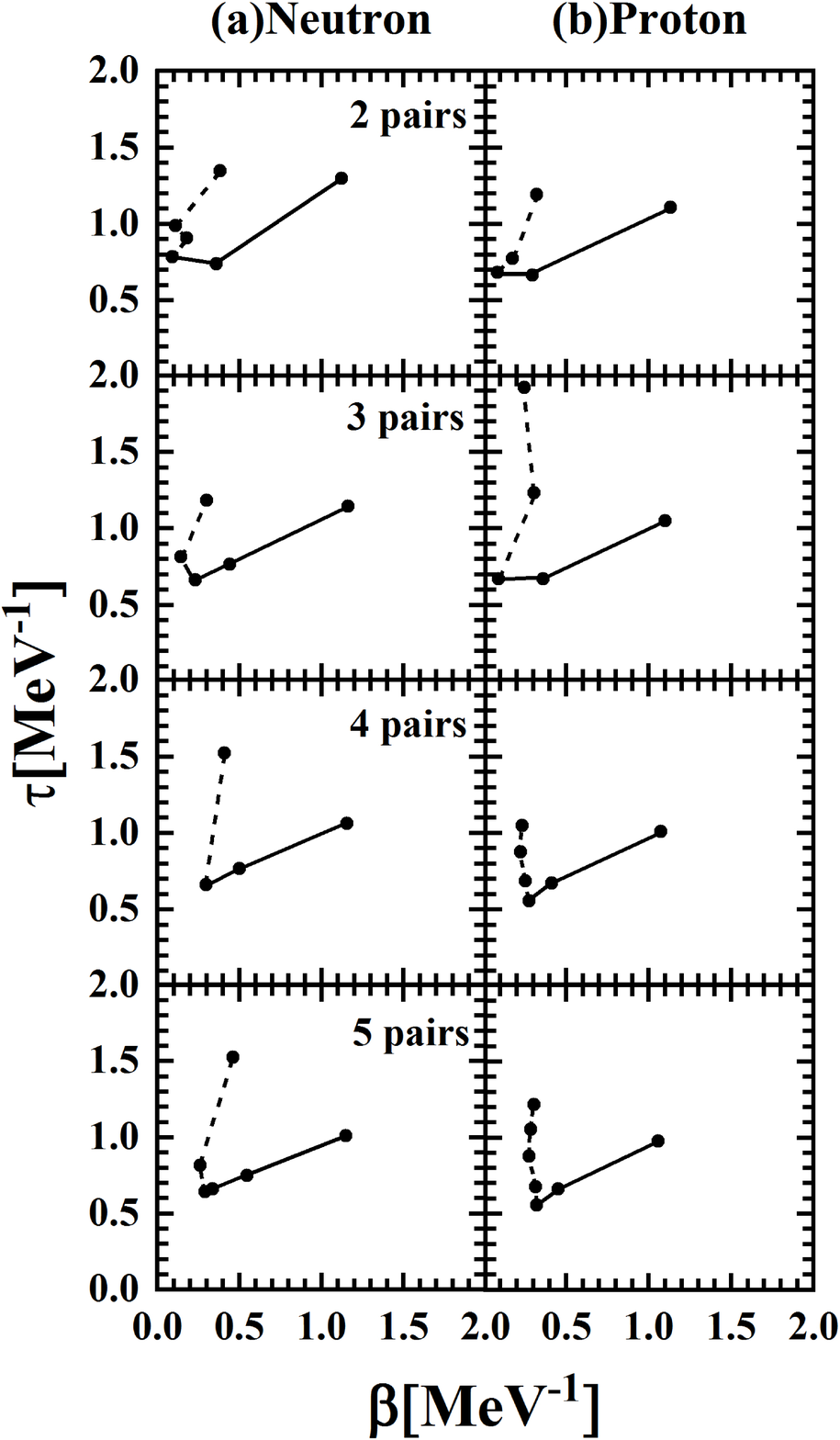}
  \caption{The DOZ of the partition function of the (a) neutron, (b) proton for $^{162} \mathrm{Dy} $ in the complex temperature plane. The Solid line and dashed line represent two kinds of phase transitions.} 
\label{fig:example4}
\end{figure}
Figure $\ref{fig:example4}$ shows the DOZ of the partition function of the (a) neutron, (b) proton for $^{162} \mathrm{Dy}$ in the complex temperature plane. The solid points correspond to the zero points in Fig.$\ref{fig:example2}$ and Fig.$\ref{fig:example3}$. The solid line and dashed line are used to distinguish two kinds of zero-point connection sequences. It can be found that with the increasing number of pairs, the regularity of the DOZ points on these two lines is gradually apparent. According to Refs.~\cite{PhysRevC.66.024322,PhysRevLett.84.3511,PhysRevA.64.013611}, these two lines can be interpreted as two different types of phase transitions. The zero points connected by the dashed line appear at small $\beta$. It means that the temperature is very high and beyond the range considered in our calculation. This can also be illustrated by the heat capacity curve in Fig.$\ref{fig:example1}$, where the heat capacity curves of neutron and proton decline at high temperature because of the limitation of our model space. Therefore, this phase transition presented in our results is unphysical.

Meanwhile, the phase transition from the superfluid phase to the normal phase illustrated by the solid line can be investigated using the modified classification scheme in Ref.~\cite{PhysRevC.66.024322}. The $\alpha$ defined in Eq.~\ref{eq:1} with 2, 3, 4, and 5 pairs of neutrons and protons have been calculated and listed in Table $\ref{tab:1}$. All the results indicate that this phase transition is a first-order phase transition.

\begin{table}[ht!]
\footnotesize
\caption{The value of $\alpha$ with 2, 3, 4, 5 pairs of neutrons and protons.}
\label{tab:1}
\doublerulesep 1pt \tabcolsep 7pt 
\begin{tabular}{ccccc}
\toprule
  & 2 pairs & 3 pairs & 4 pairs & 5 pairs \\\hline
  neutron & $\alpha$=-4.61 & $\alpha$=-4.31 & $\alpha$=-4.63 & $\alpha$=-4.74 \\
  proton & $\alpha$=-5.5 & $\alpha$=-4.57 & $\alpha$=-4.55 & $\alpha$=-4.73  \\
\toprule
\end{tabular}

\end{table}



\begin{figure}[ht]
\centering
    \includegraphics[scale=0.13]{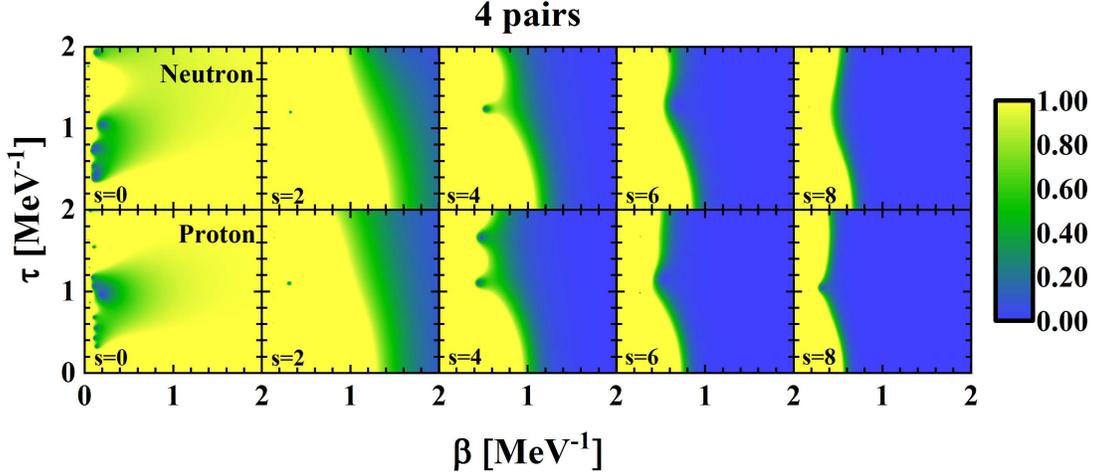}
    \caption{(Color online) The contour plots of the partition function  with different seniority numbers $s$ = 0,2,4,6,8 with four pairs of neutrons (top row) and protons (bottom row) for $^{162} \mathrm{Dy} $ in the complex temperature plane.} 
\label{fig:example5}
\end{figure}
Figure $\ref{fig:example5}$ shows the contour plots of the partition function with different seniority numbers $s$ = 0,2,4,6,8 with four pairs of neutrons (top row) and protons (bottom row) for $^{162}\mathrm{Dy} $ in the complex temperature plane. Each column represents different seniority numbers. In the first column ($s$ = 0), the zero points appear only at small $\beta$ and correspond to the unphysical phase transition discussed above. The rest non-zero regions in the complex temperature plane mean the nucleus is in the superfluid phase.

It can be found that the zero region starts to appear from the $s$ = 2 states. The normal and superfluid phases coexist in the complex temperature plane, and the boundary is obvious. As the seniority numbers increase from $s$ = 2 to 8, the normal phase dominates gradually. In the $s$ = 8 (all pairs broken) states, the entire complex temperature plane is filled by zeros except for some regions at small $\beta$, representing the nucleus is entirely in the normal phase.

\begin{figure}[h]
\centering
 \includegraphics[scale=0.1]{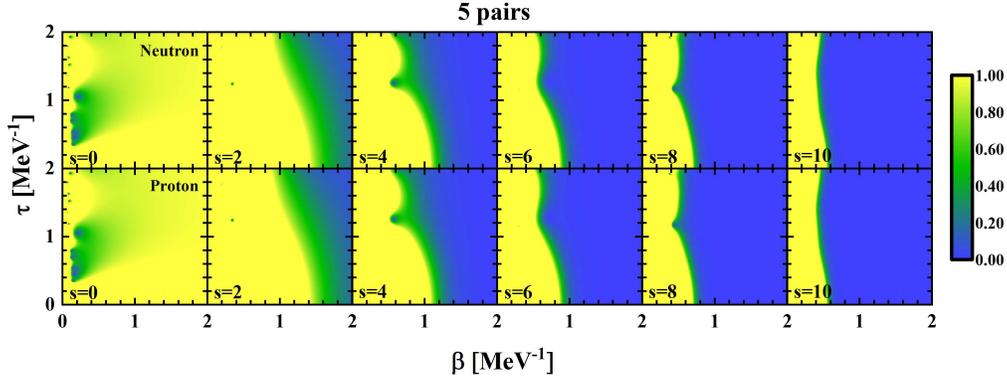}
    \caption{(Color online) The contour plots of the partition function  with different seniority numbers $s$ = 0,2,4,6,8,10 with five pairs of neutrons (top row) and protons (bottom row) for $^{162} \mathrm{Dy} $ in the complex temperature plane.} 
\label{fig:example6}
\end{figure}

The contour plots of the partition function with five pairs of particles of neutrons (top row) and protons (bottom row) for $^{162} \mathrm{Dy} $ with pairing in the complex temperature plane with different seniority numbers $s$ = 0,2,4,6,8,10 are shown in Fig.$\ref{fig:example6}$. Similar to Fig.$\ref{fig:example5}$, as the seniority numbers increase, it shows the evolution of the phase transition of pairing correlations from the superfluid state to the normal state.

\section{Summary}
\label{sec:4}
In summary, the phase transition of pairing correlations in hot nuclei $^{162} \mathrm{Dy}$ has been investigated by the CDFT+SLAP method. The order of the phase transition has been distinguished by the classification scheme based on the distribution of zeros of the pairing partition function in the complex temperature plane.

The thermodynamic quantities have been evaluated in the canonical ensemble theory. An apparent S-shaped heat capacity concerning the real temperature has been shown and indicates the phase transition from the superfluid phase to the normal phase. Moreover, the partition function calculation is extended to the complex temperature plane. Significant differences between the results with and without pairing indicate whether the pairing phase transition could happen. The DOZ of the partition function with pairing can be connected by two kinds of lines which represent two kinds of phase transition. One appears at small $\beta$ and means this phase transition could be unphysical because it goes beyond the range considered in our calculation. The other represents the transition from the superfluid phase to the normal phase. Here, one calculates the negative values of $\alpha$ for this phase transition, indicating a first-order phase transition. Besides, the contour plots of the partition function with different seniority numbers have been shown. In $s$ = 0 states, the non-zero regions fill the whole complex temperature plane (except some small $\beta$ regions), which indicates that the nucleus is in the superfluid phase. As the seniority numbers increase, the zero regions appear and gradually fill the complex temperature plane. It means the normal phase coexists with the superfluid phase and becomes dominant gradually. In all-pairs-broken states, the zero regions fill the whole complex temperature plane (except some small $\beta$ regions), which indicates that the nucleus is in the normal phase.

\begin{acknowledgments}
This work was supported by National Natural Science Foundation of China under Grant No. 11775099.
\end{acknowledgments}


\bibliographystyle{apsrev4-1}
\bibliography{PRC}



\end{document}